\documentclass{PoS} 

\title{Critical properties of $3D$ $Z(N)$ lattice gauge theories at finite 
temperature}

\ShortTitle{Critical properties of $3D$ $Z(N)$ LGTs at finite temperature}

\author{Oleg Borisenko\\
        BITP, National Academy of Sciences of Ukraine,
        03680 Kiev, Ukraine \\
        E-mail: \email{oleg@bitp.kiev.ua}}

\author{Volodymyr Chelnokov\\
        BITP, National Academy of Sciences of Ukraine,
        03680 Kiev, Ukraine\\
        E-mail: \email{chelnokov@bitp.kiev.ua}}

\author{Gennaro Cortese\\
        Instituto de F\'isica Te\'orica UAM/CSIC, Cantoblanco, E-28049 Madrid, 
        Spain,\\
        \& Departamento de F\'isica Te\'orica, Universidad de Zaragoza, 
        E-50009 Zaragoza, Spain\\
        E-mail: \email{cortese@unizar.es}}

\author{Mario Gravina\\
        Department of Physics, University of Cyprus, P.O. Box 20357, Nicosia, 
        Cyprus\\
        E-mail: \email{gravina@ucy.ac.cy}}

\author{\speaker{Alessandro Papa}\\
        Dipartimento di Fisica, Universit\`a della Calabria,\\
        \& INFN - Gruppo Collegato di Cosenza, I-87036 Rende, Italy\\
        E-mail: \email{papa@cs.infn.it}}

\author{Ivan Surzhikov\\
        BITP, National Academy of Sciences of Ukraine,
        03680 Kiev, Ukraine \\
        E-mail: \email{i\_van\_go@inbox.ru}}

\abstract{The phase structure of three-dimensional $Z(N>4)$ lattice gauge 
theories at finite temperature is investigated. Using the dual formulation of 
the models and a cluster algorithm we locate the critical points of the two 
transitions, determine various critical indices and compute average action and 
specific heat. Results are consistent with two transitions of infinite order, 
belonging to the universality class of two-dimensional $Z(N)$ vector spin 
models.}

\FullConference{31st International Symposium on Lattice Field Theory - LATTICE 2013\\
		July 29 - August 3, 2013\\
		Mainz, Germany}

\begin{document}

\section{Introduction}

$Z(N)$ lattice gauge theories (LGTs), at $T$ = 0 and $T > 0$, in addition to
being interesting on their own, can provide for useful insights into the
universal properties of $SU(N)$ LGTs, being $Z(N)$ the center subgroup
of $SU(N)$. The most general action for the $Z(N)$ LGT can be written as 
\begin{equation} 
S_{\rm gauge} \ = \ \sum_x \sum_{n<m} \ \sum_{k=1}^N \beta_k
\cos \left( \frac{2 \pi k}{N} \left(s_n(x) + s_m(x+e_n) 
-s_n(x+e_m) - s_m(x) \right) \right) \ .
\label{action_gauge}
\end{equation}
Gauge fields are defined on links of the lattice and take on values 
$s_n(x)=0,1,\cdots,N-1$. $Z(N)$ gauge models, similarly to their spin cousins, 
can generally be divided into two classes - the standard Potts models and the 
vector models. 
The standard gauge Potts model corresponds to the choice when all 
$\beta_k$ are equal. Then, the sum over $k$ in~(\ref{action_gauge}) reduces 
to a delta-function on the $Z(N)$ group. The conventional vector model 
corresponds to $\beta_k=0$ for all $k>1$. For $N=2,3$ the Potts and vector 
models are equivalent. 

While the phase structure at $T=0$ of the general model defined 
by~(\ref{action_gauge}) remains unknown, it is well established that the 
Potts models and vector models with only $\beta_1\neq 0$ have one 
phase transition from a confining phase to a phase with vanishing string 
tension~\cite{horn,ukawa}. 
Via duality, $Z(N)$ gauge models can be exactly related to $3D$ $Z(N)$ spin 
models. In particular, a Potts gauge theory is mapped to a Potts spin model, 
and such a relation allows to establish the order of the phase 
transition. Hence, Potts LGTs with $N=2$ have a second order phase transition, 
with $N\geq 3$ a first order phase transition. 
Vector models have been studied numerically in~\cite{bhanot} up to $N=20$;
for $T=0$ they exhibit a single phase transition which disappears for 
$N\to\infty$; however, their critical behavior has never been studied in 
detail.

The deconfinement phase transition at $T>0$ is well understood 
and studied for $N=2,3$. An especially detailed study~\cite{3D_z2_potts}
was performed on the gauge Ising model, $N=2$.
These models belong to the universality class of $2D$ $Z(N)$ spin models and 
exhibit a second order phase transition in agreement with the Svetitsky-Yaffe 
conjecture~\cite{svetitsky}. 
One should expect on general grounds that the gauge Potts models possess a 
first order phase transition for all $N> 4$, similarly to $2D$ Potts models. 
The $Z(4)$ vector model has been simulated, {\it e.g.}, in~\cite{3D_z4_at}. It 
also belongs to the universality class of the $2D$ $Z(4)$ spin model and 
exhibits a second order transition. Much less is known about the 
finite-temperature deconfinement transition for the vector $Z(N)$ LGTs when 
$N>4$.

In recent papers~\cite{3d_zn_strcoupl,lat_12} we considered
the vector $Z(N)$ LGTs for $N>4$ on an anisotropic lattice in the limit where 
the spatial coupling vanishes. In this limit the spatial gauge fields can be 
exactly integrated out and one gets a $2D$ generalized $Z(N)$ model, with
the Polyakov loops playing the role of $Z(N)$ spins. We found that
(i) the model shows two Berezinskii-Kosterlitz-Thouless (BKT)~\cite{berezin}
phase transitions~\footnote{For further examples of manifestation of the 
BKT transition, we refer the reader to Refs.~\cite{BKT_us}, where 
numerical techniques similar to those considered here have been adopted.}, 
(ii) for $\beta<\beta_{\rm c}^{(1)}$, there is a low-temperature, confining 
phase, with non-zero string tension $\sigma$ and linear potential,
(iii) for $\beta_{\rm c}^{(1)}<\beta<\beta_{\rm c}^{(2)}$, there is 
an intermediate phase, where the $Z(N)$ symmetry is enhanced to $U(1)$ 
symmetry, the string tension vanishes and the potential is logarithmic 
(confining),
(iv) for $\beta_{\rm c}^{(2)}<\beta$, there is a high-temperature, deconfining 
phase, with spontaneous breaking of the $Z(N)$ symmetry,
(v) critical indices are as in $2D$ vector spin $Z(N)$ models, {\it i.e.}
$\eta(\beta^{(1)}_{\rm c})=1/4$ and $\nu=1/2$ at the first transition point, 
as in the $2D$ $XY$ model, while $\eta(\beta^{(2)}_{\rm c})=4/N^2$ and 
$\nu=1/2$ at the second transition point.

The aim of this work is to extend the analysis to $3D$ vector $Z(N>4)$ LGTs 
at $T>0$ on isotropic lattices with $\beta_s=\beta_t\equiv \beta$. If, as 
probable, spatial plaquettes have small influence on the dynamics of 
the Polyakov loop interaction, we expect the same scenario as in the
model with $\beta_s=0$.

\section{From the $3D$ $Z(N)$ LGT to a generalized $3D$ $Z(N)$ spin
model}  

We work on a $3D$ lattice $\Lambda = L^2\times N_t$ with spatial extension $L$ 
and temporal extension $N_t$; $\vec{x}=(x_0,x_1,x_2)$, where $x_0\in [0,N_t-1]$
and $x_1,x_2\in [0,L-1]$ denote the sites of the lattice and $e_n$, $n=0,1,2$, 
denotes a unit vector in the $n$-th direction.
Periodic boundary conditions on gauge fields are imposed in all 
directions. The notations $p_t$ ($p_s$) stand for the temporal (spatial) 
plaquettes, $l_t$ ($l_s$) for the temporal (spatial) links.
We introduce conventional plaquette angles $s(p)$ as
\begin{equation}
s(p) \ = \ s_n(x) + s_m(x+e_n) - s_n(x+e_m) - s_m(x) \ .
\label{plaqangle}
\end{equation}
The $3D$ $Z(N)$ gauge theory on an anisotropic lattice can generally be 
defined as 
\begin{equation}
Z(\Lambda ;\beta_t,\beta_s;N) \ = \  \prod_{l\in \Lambda}
\left ( \frac{1}{N} \sum_{s(l)=0}^{N-1} \right ) \ \prod_{p_s} Q(s(p_s)) \
\prod_{p_t} Q(s(p_t)) \ .
\label{PTdef}
\end{equation}
The most general $Z(N)$-invariant Boltzmann weight with $N-1$ different 
couplings is
\begin{equation}
Q(s) \ = \
\exp \left [ \sum_{k=1}^{N-1} \beta_p(k) \cos\frac{2\pi k}{N}s \right ] \ .
\label{Qpgen}
\end{equation}
The Wilson action corresponds to the choice $\beta_p(1)=\beta_p$, 
$\beta_p(k)=0, k=2,...,N-1$. By standard duality transformation 
(see, {\it e.g.},~\cite{ukawa}), one gets a generalized $3D$ spin $Z(N)$ 
model, with action
\[
S \ =\ \sum_{x}\ \sum_{n=1}^3 \sum_{k = 1}^{N-1} \ \beta_k \  
\cos \left( \frac{2 \pi k}{N} \left(s(x) - s(x+e_n) \right) \right)\;,
\;\;\;
\beta_k \ =\ \frac{1}{N} \sum_{p = 0}^{N - 1} \ln \left [ \frac{Q_d(p)}
{Q_d(0)} \right ] \  \cos \left(\frac{2 \pi p k}{N} \right) \ .
\]
It can be shown that the dual model is ferromagnetic and that, generally,
$| \beta_1 | \gg | \beta_2 |$. Thus, one expects that the $3D$ vector spin 
model with only $\beta_1\neq 0$ gives a reasonable approximation to 
the gauge model (in our simulations we use all $\beta_k$). Next important 
fact, is that the weak and the strong coupling regimes are interchanged: 
when $\beta\to\infty$ the effective couplings $\beta_k\to 0$ and, therefore, 
the ordered symmetry-broken phase is mapped to a symmetric phase with 
vanishing magnetization of dual spins. The symmetric phase at small 
$\beta$ becomes an ordered phase where the dual magnetization is 
non-zero (see~\cite{3dzn} for details).

\section{Numerical results}

The BKT transition, being of infinite order, is hard to study by
analytical methods, such as renormalization group technique 
of Ref.~\cite{elitzur}. Numerical simulations are plagued by the 
very slow, logarithmic convergence to the thermodynamic limit in the vicinity 
of the BKT transition, thus calling for large-scale simulations in combination 
with finite-size scaling methods.

The standard approach would consist in the using Binder cumulants to locate 
the position of critical points and susceptibilities in order to determine the
critical indices. Both Binder cumulants and susceptibilities should be 
constructed from Polyakov loops, but the expression of a single Polyakov loop 
is non-trivial in the dual formulation.

Here we follow a different strategy, consisting in the use of Binder 
cumulants and susceptibilities constructed {\em from the dual $Z(N)$ spins}.
Interestingly, the critical behavior of dual spins is reversed with respect 
to the critical behavior of Polyakov loops: (i) the spontaneously-broken 
ordered phase is mapped to the symmetric phase and {\it vice versa} and the 
critical indices $\eta$ are also interchanged, (ii) the index $\nu$ which 
governs the exponential divergence of the correlation length is expected to 
be the same at both transitions and takes on the value $\nu=1/2$ 
(see~\cite{3dzn} for details).

We simulate the $3D$ $Z(N)$ dual model by a cluster algorithm, 
with all the couplings $\beta_k$, for $N=5,\ 8,\ 13,\ 20$, on an 
$N_t\times L^2$ lattice with periodic boundaries, 
with $N_t=2,\ 4,\ 6,\ 8,\ 12$ ~\footnote{In Ref.~\cite{3dzn} we have given 
results of simulations for $N$=5 and 13 and $N_t$=2 and 4. The results for
other values of $N$ and $N_t$ are new and a paper is in 
preparation~\cite{prep}, where also the continuum limit and scaling with $N$ 
will be investigated.}. 
The typical statistics is $10^6$
(equilibration after $10^5$ configurations, measurements taken every 
10 updating steps; error analysis by jackknife combined with binning). 
The adopted observables are
\begin{itemize}
\item complex magnetization $M_L = |M_L| e^{i \psi}$, \ 
$ M_L \ =\  \sum_{x \in \Lambda} \exp \left( \frac{2 \pi i}{N} s(x) \right)$

\item population \ $S_L \ =\  \frac{N}{N - 1} \left(\frac{\max_{i = 0, N - 1} n_i} {L^2 N_t}- \frac{1}{N} \right)$,
where $n_i$ is number of $s(x)$ equal to $i$

\item real part of the rotated magnetization $M_R = |M_L| \cos(N \psi)$
and normalized rotated magnetization $m_\psi = \cos(N \psi)$

\item susceptibilities of $M_L$, $S_L$ and $M_R$: 
$\chi_L^{(M)}$, $\chi_L^{(S)}$, $\chi_L^{(M_R)}$, where
$\chi_L^{(\mathbf\cdot)} \ =\  L^2 N_t \left(\left< {\mathbf\cdot}^2 \right> 
- \left< \mathbf\cdot \right>^2 \right)$

\item Binder cumulants 
$U_L^{(M)}\ =\ 1 - \frac{\left\langle \left| M_L \right| ^ 4 
\right\rangle}{3 \left\langle \left| M_L \right| ^ 2 \right\rangle^2}$
and
$B_4^{(M_R)}\ =\  \frac{\left\langle \left| M_R 
- \left\langle M_R \right\rangle \right| ^ 4 \right\rangle}
{\left\langle \left| M_R - \left\langle M_R \right\rangle \right| ^ 2 
\right\rangle ^ 2 }\;$.

\end{itemize}

\begin{figure}[tb]
\centering
\includegraphics[width=0.28\textwidth]{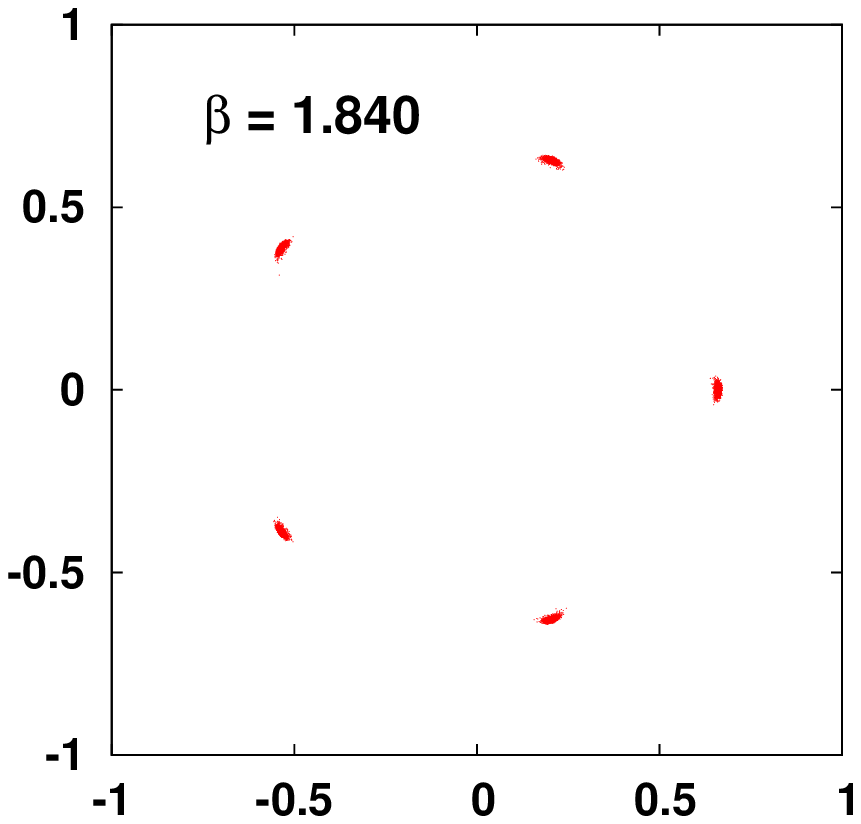}
\includegraphics[width=0.28\textwidth]{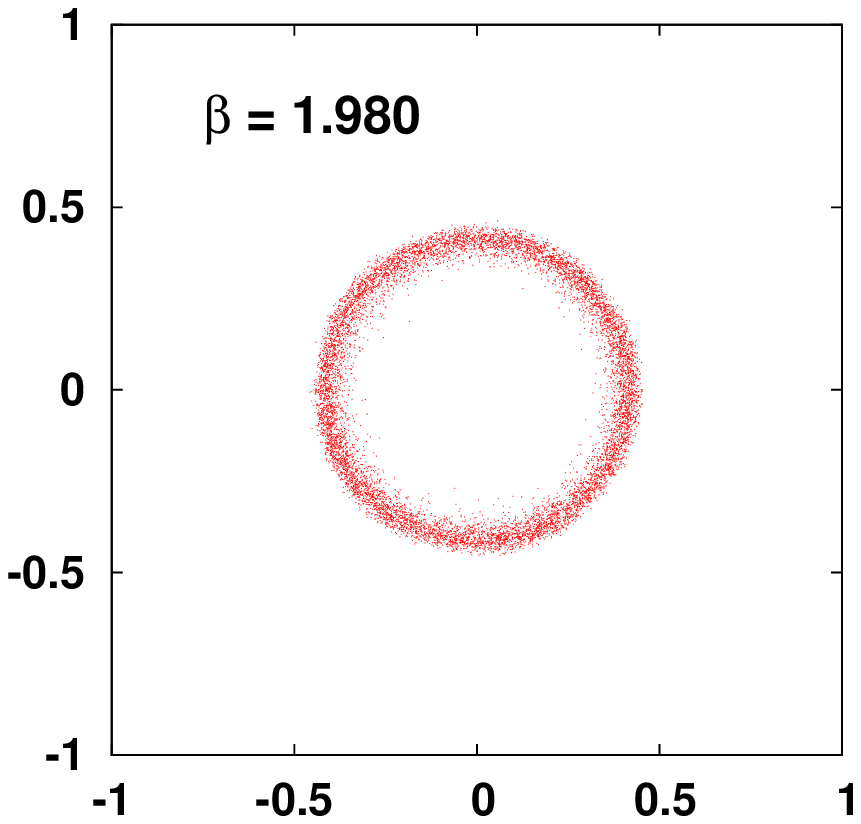}
\includegraphics[width=0.28\textwidth]{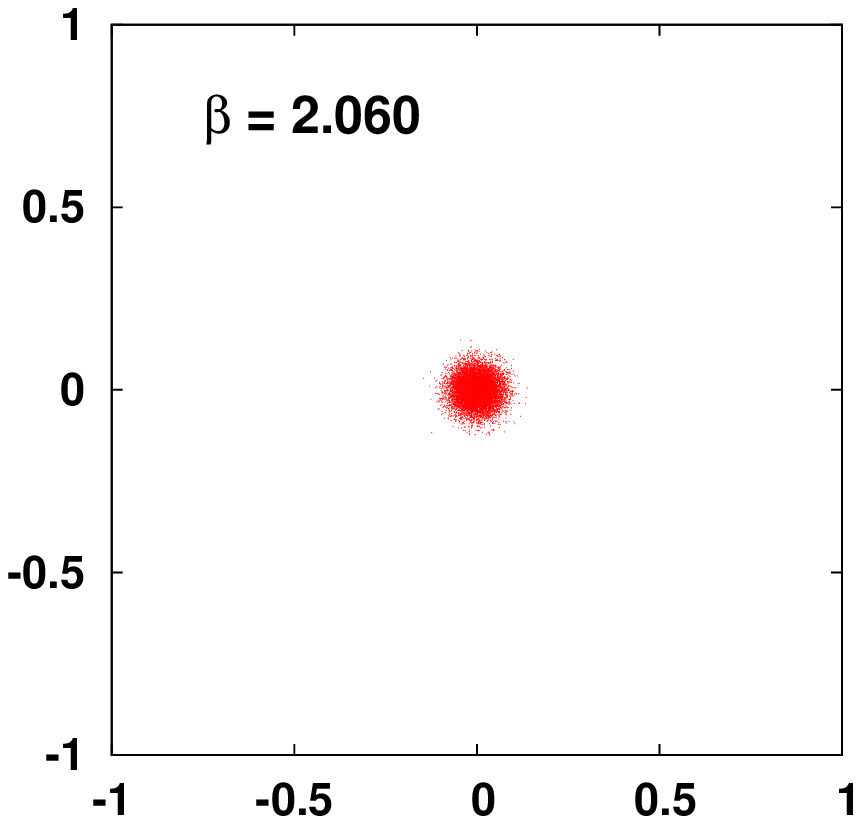}
\caption{Scatter plot of $M_L$ at $\beta$=1.84, 1.98 and 2.06 in $Z(5)$ 
on a $512^2\times 4$ lattice.}
\label{fig:scatter}
\end{figure}

A clear indication of the three-phase structure emerges from the inspection
of the scatter plot of the complex magnetization $M_L$ at different values 
of $\beta$: as we move from low to high $\beta$, we observe the transition
from an ordered phase ($N$ isolated spots)  through an intermediate phase 
(ring distribution) up to the disordered phase (uniform distribution around 
zero) -- see Fig.~\ref{fig:scatter}.

\begin{figure}[tb]
\includegraphics[width=0.49\textwidth]{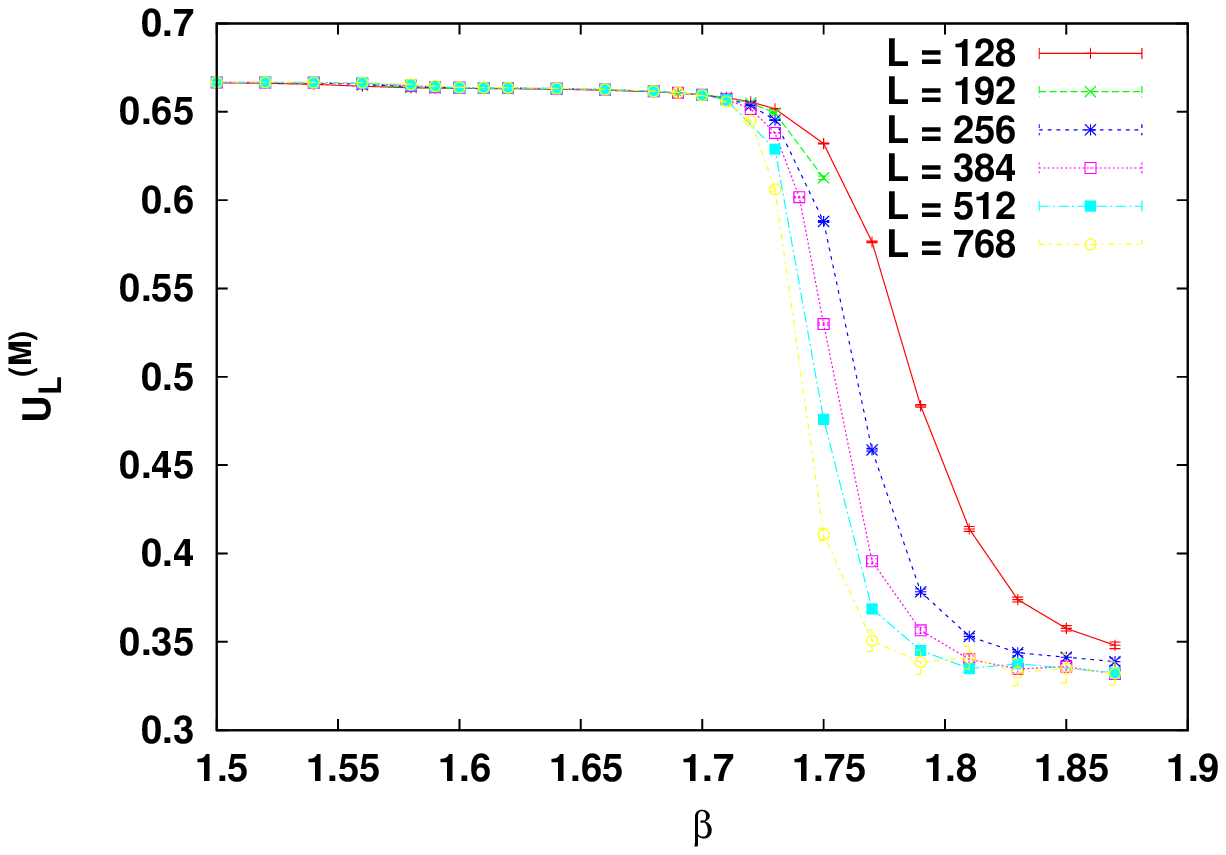}
\includegraphics[width=0.49\textwidth]{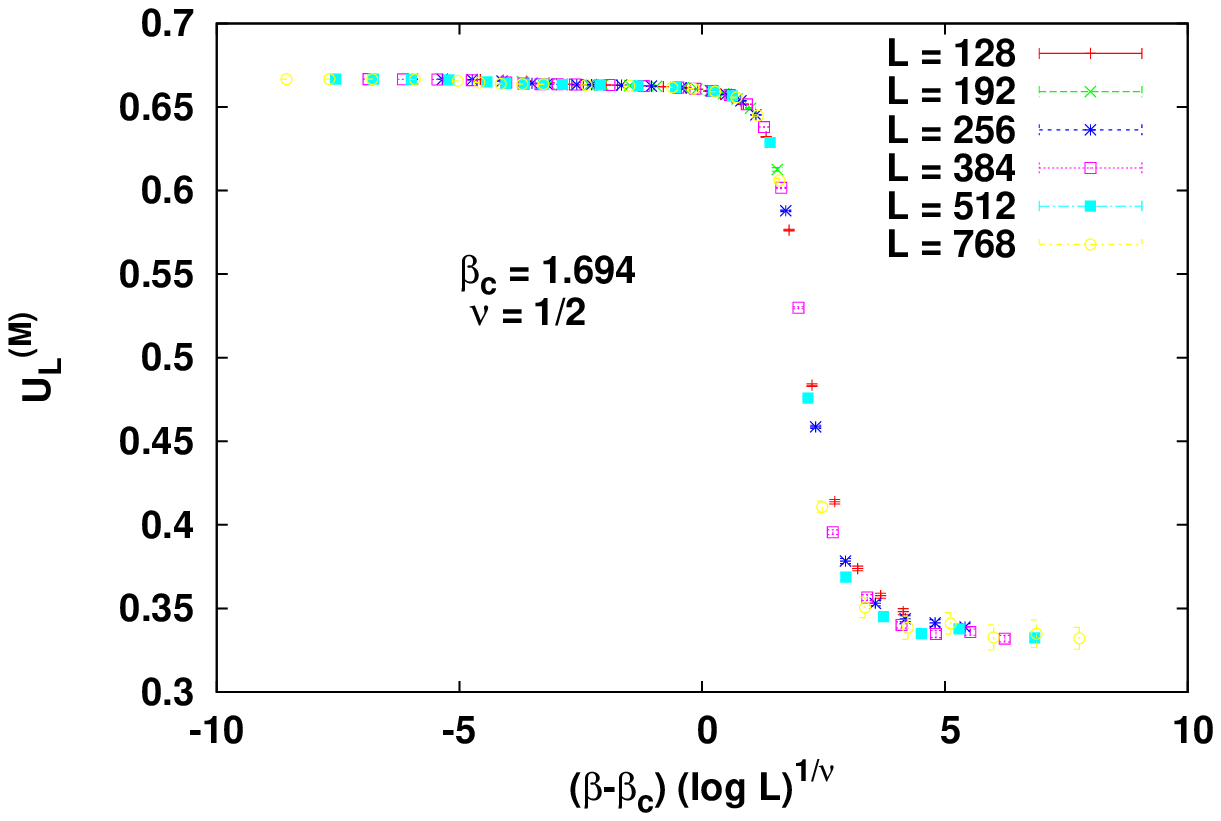} 
\caption{$U_L^{(M)}$ as function of $\beta$ (left) and
 of $(\beta-\beta_c) (\ln L)^{1/\nu}$ (right) in $Z(5)$ $N_t = 2$ model.}
\label{U_plots_Z5_NT2}
\end{figure}

The first step is to determine the two critical couplings in the 
thermodynamic limit, $\beta_{\rm c}^{(1)}$ and $\beta_{\rm c}^{(2)}$, that 
separate the three phases. To this aim we find the value of $\beta_c$ which 
provides the best overlap of universal observables, plotted for different 
values of $L$ against $(\beta-\beta_{\rm c}^{(1)})(\ln L)^{1/\nu}$, 
with $\nu$ fixed at 1/2. As universal observables we used the 
Binder cumulant $B_4^{(M_R)}$ and the order parameter $m_{\psi}$ for 
the first phase transition and the Binder cumulant $U_L^{(M)}$ for the second 
phase transition.
In Fig.~\ref{U_plots_Z5_NT2} we show as an example the plots of one of these
universal observables against $\beta$ and against 
$(\beta-\beta_{\rm c}^{(1)})(\ln L)^{1/\nu}$, with $\nu$ fixed at 1/2. 
In Table~\ref{tbl:crit_betas} we report the determinations of the critical 
couplings $\beta_{\rm c}^{(1)}$ and $\beta_{\rm c}^{(2)}$, in $Z(N)$ 
with $N$=5, 8, 13 and 20, for $N_t$=2, 4, 8 and 12.

\begin{table}[tb]
\caption{Values of $\beta_{\rm c}^{(1)}$ and $\beta_{\rm c}^{(2)}$ obtained 
for various $N_t$ in $Z(N)$ with $N = 5,\ 8,\ 13$ and 20.}
\begin{center}
\begin{tabular}{|c|c|c|c|}
\hline
$N$ & $N_t$ & $\beta_{\rm c}^{(1)}$ & $\beta_{\rm c}^{(2)}$ \\
\hline
 5 & 2 & 1.617(2) & 1.694(2) \\
 5 & 4 & 1.943(2) & 1.990(2) \\ 
 5 & 6 & 2.05(1)  & 2.08(1) \\
 5 & 8 & 2.085(2) & 2.117(2) \\
 5 & 12 & 2.14(1) & 2.16(1) \\
\hline
 8 & 4 & 2.544(8) & 4.688(5) \\
 8 & 8 & 3.422(9) & 4.973(3)\\
\hline
\end{tabular}
\hspace{1cm}
\begin{tabular}{|c|c|c|c|}
\hline
$N$ & $N_t$ & $\beta_{\rm c}^{(1)}$ & $\beta_{\rm c}^{(2)}$ \\
\hline
 13 & 2 & 1.795(4) & 9.699(6) \\
 13 & 4 & 2.74(5)  & 11.966(7) \\
 13 & 8 & 3.358(7) & 12.710(2) \\
\hline
 20 & 4 & 2.57(1) & 28.15(2)\\
 20 & 8 & 3.42(5) & 29.731(4)\\
\hline
\end{tabular}
\end{center}
\label{tbl:crit_betas}
\end{table}

Now, we are able to extract some critical indices and check the hyperscaling 
relation. Since we are using the observables in the dual model, the 
transitions change places: the first transition is governed by the behavior of 
$M_{R}$, the second one by the behavior of $M_{L}$.

We start the discussion from the second transition. According to the 
standard finite-size scaling (FSS) theory, the equilibrium magnetization 
$|M_{L}|$ at criticality should obey the relation 
$|M_{L}| \sim L^{-\beta / \nu}$, if the spatial extension $L$ of the lattice
is large enough. Therefore, we fit data of $|M_L|$ at $\beta^{(2)}_{\rm c}$, 
on all lattices with size $L$ not smaller than a given $L_{\rm min}$, with the 
scaling law $|M_{L}|=A L^{-\beta/\nu}$.
The FSS behavior of the susceptibility $\chi^{(M)}_L$ is given by 
$\chi^{(M)}_L\sim L^{\gamma/ \nu}$, where $\gamma/\nu=2-\eta$ and $\eta$ is 
the magnetic critical index. Therefore we fit data of $\chi^{(M)}_L$ at
$\beta^{(2)}_{\rm c}$, on all lattices with size $L$ not smaller 
than a given $L_{\rm min}$, according to the scaling law 
$\chi^{(M)}_{L}= A L^{\gamma/\nu}$.

The reference value for the index $\eta$ at this transition is 1/4, whereas the
the hyperscaling relation to be fulfilled is $\gamma/\nu+2\beta/\nu=d=2$.
We find (see~\cite{3dzn,prep} for details) that in most cases the values of 
$\eta$ and $d$ are close to those predicted by universality. The small
discrepancy from the exact values $\eta = 0.25$ and $d = 2$ may be caused by 
the asymptotically vanishing parts of the scaling behavior of the observables 
$|M_L|$ and $\chi^{(M)}_L$, that we are not taking into account, but may 
be significant for smaller lattice sizes.

The procedure for the determination of the critical indices at the first
transition is similar to the one for the second transition, with the difference
that the scaling laws given above are to be applied to the rotated 
magnetization, $M_R$, and to its susceptibility, $\chi_L^{(M_R)}$, 
respectively.

The reference value for the index $\eta$ at this transition is 
$4/N^2$, {\it i.e.} $\eta=0.16$ for $N=5$ and $\eta\approx 0.0237$ for $N=13$,
whereas the hyperscaling relation to be fulfilled is $\gamma/\nu+2\beta/\nu=d
=2$. Also here we have found (see~\cite{3dzn,prep} for details) a general 
agreement between the $\eta$ and $d$ values obtained and those predicted by 
universality. However, the expected value of $\beta/\nu$ is very 
small, ($2/N^2$), so other, asymptotically vanishing, terms can have a great 
impact on its determination on finite-sized lattices. This is especially 
evident for $Z(13)$ with $N_t=4$, where $\beta/\nu$ turned out to be negative 
indicating that the magnetization $M_R$ grows with lattice size. 

\begin{figure}
\includegraphics[width=0.49\textwidth]{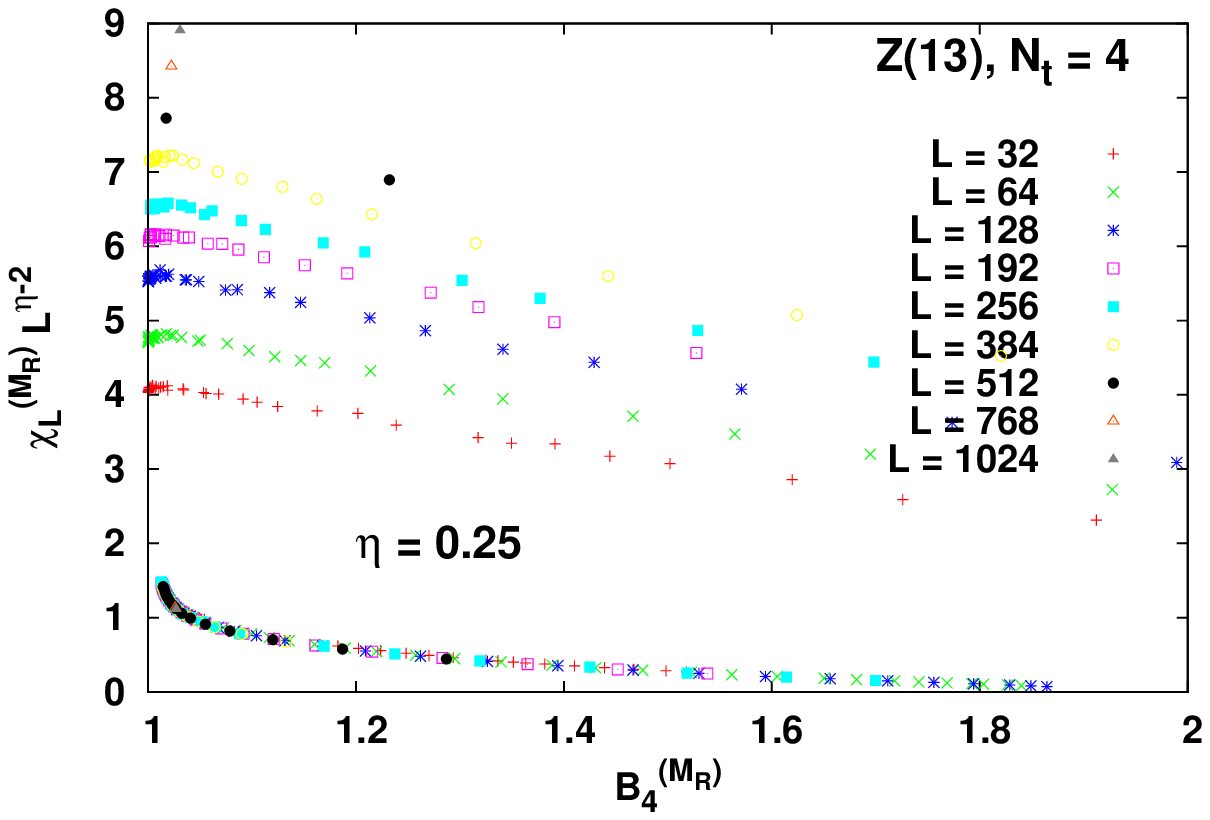}
\includegraphics[width=0.49\textwidth]{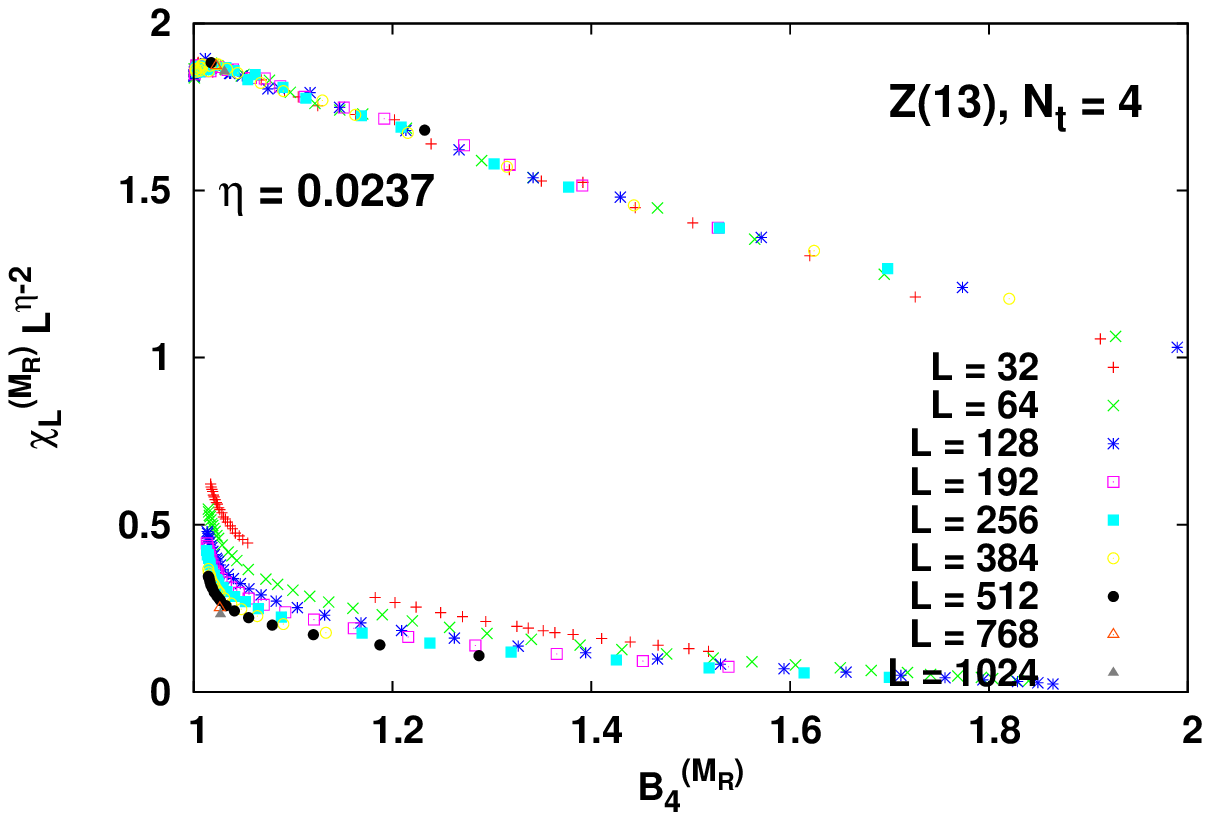}  
\caption{Correlation between $\chi_{L}^{(M_{R})}L^{\eta-2}$ and 
the Binder cumulant $B_{4}^{(M_{R})}$ in $Z(13)$ with $N_t=4$ for $\eta=0.25$ 
(left) and for $\eta=0.0237$ (right) on lattices with different size.}
\label{chimr_vs_b_Z13_NT4}
\end{figure}

There is an independent method to determine the critical exponent $\eta$,
which does not rely on the prior knowledge of the critical coupling, but 
is based on the construction of a suitable universal 
quantity~\cite{Loison99,2dzn}. The idea is to plot 
$\chi_{L}^{(M_{R})}L^{\eta-2}$ versus $B_{4}^{(M_{R})}$ and to look for 
the value of $\eta$ which optimizes the overlap of curves from different 
volumes. This method is illustrated in Fig~\ref{chimr_vs_b_Z13_NT4}. 
for $Z(13)$ model with $N_t = 4$.

Concerning the value of the critical index $\nu$, the methods used in this work
do not allow for the direct determination of its value. When locating critical 
points we have fixed $\nu$ at 1/2. This value appears to be well in agreement 
with all numerical data. 

To provide further evidence on the nature of the phase transitions we have
performed Monte Carlo simulation of the {\em original} gauge model,
in particular we considered the $Z(5)$ LGT with $N_t=2,4$ and spatial 
extent $L\in [64-512]$. The typical statistics was $10^5$.
In general, error bars are larger and results for critical indices are not 
so precise as in dual model simulations. Nevertheless, we can state 
that (i) the critical index $\eta$ is compatible with its $2D$ value;
(ii) the values of the indices at two transitions are indeed interchanged 
as explained before (see~\cite{3dzn} for details).

Finally, we have calculated the average action and the specific heat 
around the transitions in $Z(5)$ LGT with $N_t$=2 and $N_t=4$. In all cases 
the dependence of these quantities on $\beta$ turned out to be continuous, 
thus ruling out first and second order transitions and being compatible
with a transition of infinite order (see~\cite{3dzn} for details).

\section{Conclusions}

We have studied $3D$ $Z(N>4)$ vector LGTs at the finite temperature, using the 
exact dual transformation to generalized $3D$ $Z(N)$ spin models and
determined the two critical couplings of $Z(N=5,8,13,20)$ 
vector LGTs and given estimates of the critical indices $\eta$ 
at both transitions.
We have observed, for the first time in these models, a scenario with 
three phases: disordered phase at small $\beta$, massless or BKT phase at 
intermediate values of $\beta$, ordered phase at larger and larger 
values of $\beta$ as $N$ increases.  
This matches perfectly with the $N\to\infty$ limit, {\it i.e.} the $3D$ 
$U(1)$ LGT, where the ordered phase is absent.

We have found that the values of the critical index $\eta$ at the two 
transitions are compatible with the theoretical expectations.
The index $\nu$ also appears to be compatible with the value $1/2$, 
in agreement with universality predictions. 
We conclude that finite-temperature $3D$ $Z(N>4)$ vector LGTs 
undergo two phase transitions of the BKT type and belong to the 
universality class of the $2D$ $Z(N)$ vector models.

\end{document}